\documentclass[journal]{IEEEtran}

\usepackage{pgfplots}
\usepackage[bookmarks,colorlinks]{hyperref}
\usepackage[linesnumbered,ruled,lined]{algorithm2e}
\usepackage{enumitem}
\usetikzlibrary{shapes.multipart,intersections}
\usepackage{cite}
\usepackage{amsmath,amssymb,amsfonts,amsthm,steinmetz}
\usepackage{mathrsfs}  
\usepackage{textcomp}
\usepackage{acronym}
\usepackage{xcolor}
\usepackage{upgreek,xspace}
\usepackage{array}
\usepackage{tikz}
\usetikzlibrary{calc}
\makeatletter
\newcommand{\gettikzxy}[3]{%
  \tikz@scan@one@point\pgfutil@firstofone#1\relax
  \edef#2{\the\pgf@x}%
  \edef#3{\the\pgf@y}%
}
\makeatother

\usepackage[draft]{todonotes}

\usepackage{esvect}

\usepackage{times}
\usepackage{bm}
\usepackage{stmaryrd}
\usepackage{babel}
\usepackage{graphics, graphicx}
\usepackage{gensymb}
\usepackage{cite}
\usepackage{enumitem}
\usepackage{url}

\newcommand{\Sym}{\operatorname{Sym}}

\begin{document}

\title{Low-Complexity Wireless Multi-Port Sensing \\by Multiplexed De-Embedding \\of an Over-the-Air Fixture}

\author{Philipp~del~Hougne,~\IEEEmembership{Member,~IEEE}
\thanks{This work was supported in part by the ANR France 2030 program (project ANR-22-PEFT-0005), the ANR PRCI program (project ANR-22-CE93-0010), the European Union's European Regional Development Fund, and the French Region of Brittany and Rennes M\'etropole through the contrats de plan \'Etat-R\'egion program (projects ``SOPHIE/STIC \& Ondes'' and ``CyMoCoD'').}
\thanks{
P.~del~Hougne is with Univ Rennes, CNRS, IETR - UMR 6164, F-35000, Rennes, France (e-mail: philipp.del-hougne@univ-rennes.fr).
}
}

\maketitle

\begin{abstract}
Wireless multi-port sensing remotely retrieves the scattering matrix of a \textit{multi}-port device under test (DUT) connected to a set of not-directly-accessible (NDA) antennas that couple over-the-air (OTA) to a set of accessible antennas. 
If (i) the OTA fixture characteristics are known, and (ii) the number of independent measurements at the accessible antennas is sufficient, the OTA fixture can be de-embedded to recover the DUT characteristics. 
In recent prior work, we solved (i) by connecting the NDA antennas to a specific known tunable load network (TLN). 
Here, we tackle (ii) by additionally using the TLN to provide measurement diversity. 
The connection between OTA fixture and TLN constitutes a programmable fixture (PF). When the DUT characteristics cannot be identified based on a single PF realization, we add measurement diversity with multiple PF realizations.
The underlying ``multiplexed de-embedding'' achieves the joint de-embedding of an ensemble of PF realizations when a single PF realization cannot be de-embedded. 
We experimentally demonstrate our concept by remotely estimating the scattering matrix of a reciprocal, non-unitary 4-port DUT (10 complex-valued unknowns) via a rich-scattering OTA fixture purely based on measurements of a single transmission coefficient between two accessible antennas across 30 different PF realizations. 
We systematically study the trade-off between the number of independent measurements at the accessible antennas and the number of PF realizations. Multiplexed de-embedding of the OTA fixture paves the path to implementing wireless multi-port sensing with low hardware complexity in areas like RFID and wireless bioelectronics.
\end{abstract}

\begin{IEEEkeywords}
Wireless sensing, backscatter modulation, multiport-network theory, mutual coupling, Virtual VNA, reverberation chamber, MIMO, RFID, over-the-air fixture, multiplexed de-embedding, measurement diversity, compressed sensing, non-linear observation.
\end{IEEEkeywords}

\section{Introduction}
\label{sec_introduction}

Antenna backscatter modulation is the backbone of diverse communications and sensing techniques. Notable examples include the Great Seal Bug~\cite{brooker2013lev}, RFID technologies, and reconfigurable intelligent surfaces (RISs) in smart radio environments. Wireless sensing can leverage digital or analog backscatter modulation~\cite{Marrocco}. In the latter case, the sensing mechanism can modulate either the structure or the load of the backscatter antenna. Correspondingly, one can distinguish between analog wireless sensing of the antenna's reflection coefficient and of the antenna's load. In both cases, generalizations to multi-port backscatter modulation, also referred to as RFID grids~\cite{Marrocco_RFID_GRID,Marrocco_RFID_GRID2}, are relevant in situations involving either multiple single-port backscatter-modulated antennas in close proximity~\cite{marrocco2008multiport,mughal2023statistical,nanni2024stackedRFID,barbot2025differential} or a single multi-port backscatter-modulated antenna system~\cite{caizzone2011multi}. 

The remote estimation of the reflection coefficient of a backscatter-modulated antenna has received considerable attention~\cite{garbacz1964determination,mayhan1994technique,pursula2008backscattering,Capstick2009,bories2010small,van2020verification,sahin2021noncontact,kruglov2023contactless}, often under simplifying assumptions such as a free-space wireless propagation environment (WPE). Some works have also considered the generalized problem of remotely estimating the reflection matrix of a multi-port backscatter-modulated system~\cite{wiesbeck1998wide,monsalve2013multiport,denicke2012application,del2024virtual,shilinkov2024antenna,del2024virtual2p0,tapie2025scalable,del2025virtual3p0,del2025virtual3p1}. A detailed comparison of these works can be found in [Sec.~II,~\cite{del2025virtual3p0}]. The most general version of these techniques was recently developed within the ``Virtual Vector Network Analyzer'' (Virtual VNA) framework~\cite{del2024virtual,del2024virtual2p0,tapie2025scalable,del2025virtual3p0,del2025virtual3p1} that offers closed-form and gradient-descent techniques, without relying on simplifying assumptions or requiring calibration-standard loads, and that is compatible with the use of arbitrarily many probing antennas as well as non-coherent detection. 
The Virtual VNA enables the full and unambiguous retrieval of the scattering parameters of a many-port device under test (DUT) without any manual reconnections. Waves can only be injected/received via a fixed subset of the DUT ports while the remaining DUT ports are terminated by a specific tunable load network (TLN). The TLN can terminate each connected DUT port with three distinct individual loads, or with coupled loads, and the TLN scattering characteristics are known. The TLN ports thus act like ``virtual'' additional VNA ports. 

Meanwhile, the remote estimation of the load of a backscatter-modulated antenna is less mature. Various works~\cite{chen2012wireless,bjorninen2011wireless,akbar2015rfid,skrobacz2024new,chen2012coupling,vena2024backscatter,del2025wireless,barbot2025general} have tackled versions of this problem, often under assumptions such as a free-space WPE, the availability of one or multiple perfectly known antenna(s), or the availability of a perfectly matched antenna. The general problem without any of these assumptions was only tackled this year, independently in~\cite{del2025wireless} and~\cite{barbot2025general}. The formulation in~\cite{barbot2025general} based on Green's equation is specific to a backscatter-modulated single-input single-output (SISO) link and can be shown to be equivalent to the correspondingly specialized version of the multi-port network formulation in~\cite{del2025wireless}. The generalized problem of remotely estimating the scattering matrix of a multi-port load network terminating a multi-port backscatter-modulated system initially received attention in~\cite{Marrocco_RFID_GRID,Marrocco_RFID_GRID2}; these works retrieved a vector whose entries are proportional to the magnitudes of the diagonal entries of the admittance matrix of the multi-port load network. This year,~\cite{del2025wireless} demonstrated a method for unambiguously retrieving the full scattering matrix of the load network (without any simplifying assumptions). 

The key insight in~\cite{del2025wireless} was to frame the problem as the de-embedding of an over-the-air (OTA) fixture. This breaks down the remote load network characterization into two steps. \textit{First}, the NDA antennas are connected to a known TLN to characterize the OTA fixture with the established Virtual VNA technique~\cite{del2025wireless}. \textit{Second}, the NDA antennas are connected to the DUT, and the OTA fixture is de-embedded from the measurement. However, the identifiability of the load network in the second step hinges on whether the number of independent scattering coefficients measured at the OTA fixture's accessible ports is sufficient, as well as on the characteristics of the OTA fixture. While it is difficult to formulate precise conditions due to the non-linear operations, a lower bound for the required number of independent scattering coefficient measurements is the number of independent unknowns in the DUT's scattering matrix. Consequently, the technique from~\cite{del2025wireless} may require a large number of probe antennas, as well as corresponding coherent wavefront generation and reception, to remotely characterize many-port load networks.

Here, we overcome this vexing hardware complexity by making use of the TLN also in the second step. We treat the connection between the OTA fixture and the TLN as a known programmable fixture (PF) in the second step. The access to different realizations of the PF can compensate a small number of independent scattering coefficients at the OTA fixture's accessible ports. In fact, measuring a single transmission coefficient across different PF realizations can be sufficient for wireless multi-port sensing, as we demonstrate experimentally below. Thereby, the identifiability of the load network can be uncoupled from the number of independent scattering coefficients at the OTA fixture's accessible ports. Consequently, we drastically reduce the hardware complexity and cost for wireless multi-port sensing, making it compatible with the sort of low-cost SISO transmission measurements based on two software-defined radios (SDRs) that are typical in the realm of RFID~\cite{vena2024backscatter}.

The remainder of this paper is organized as follows. In Sec.~\ref{sec_ProblemStatement}, we formalize our problem statement. In Sec.~\ref{sec_CScomparison}, we clarify how our approach differs from conventional compressive sensing. In Sec.~\ref{sec_Method}, we detail our method. In Sec.~\ref{sec_ExpVal}, we present the experimental validation of our method; we describe our experimental setup in Sec.~\ref{subsec_expSetupProc}, we analyze the PF diversity in Sec.~\ref{subsec_AnalysisPFdiversity}, and we examine our estimates of the DUT characteristics in Sec.~\ref{subsec_expResults}. We conclude in Sec.~\ref{sec_conclusion} with a brief summary and outlook to future research directions.

\section{Problem Statement}
\label{sec_ProblemStatement}

\begin{figure*}
    \centering
    \includegraphics[width=2\columnwidth]{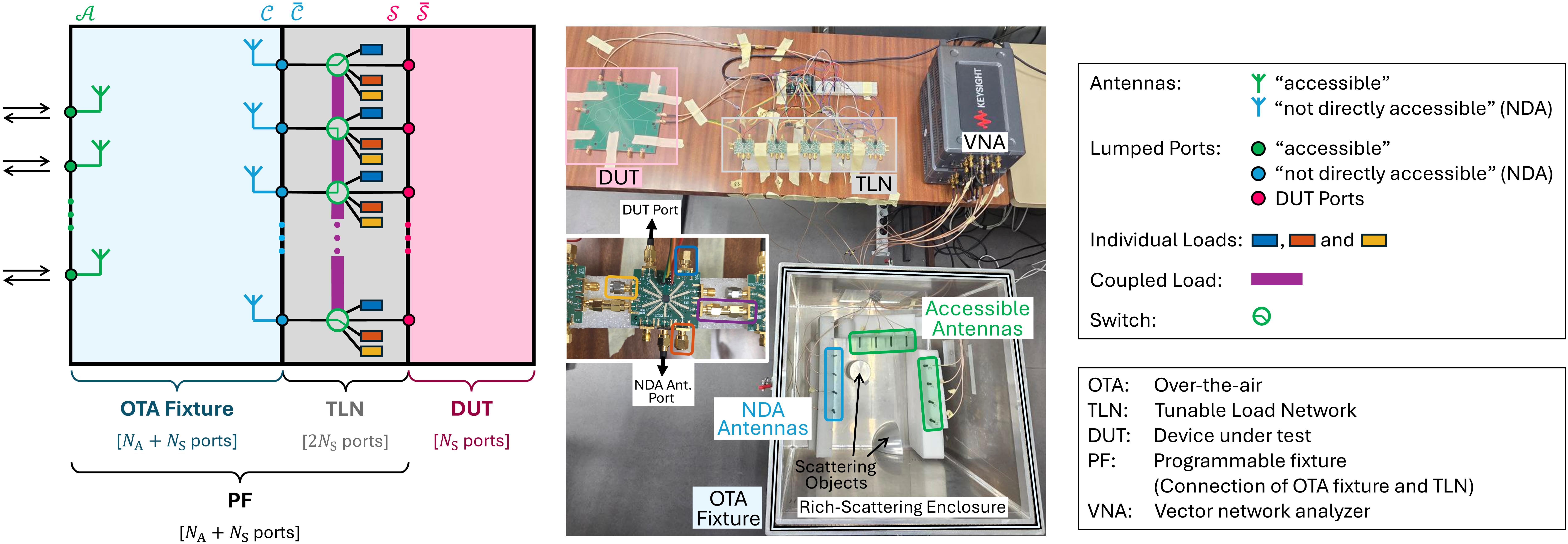}
    \caption{Our system model (left) partitions our setup (middle) into a chain cascade of three multi-port systems: the OTA fixture (comprising the rich-scattering environment surrounding the antennas as well as the antennas' structural scattering), the tunable load network (TLN), and the DUT. The cascade of OTA fixture and TLN constitutes a programmable fixture (PF) via which we probe the DUT. The relevant port index sets are indicated above the connections in the system model on the left; their cardinalities are $|\mathcal{A}|=N_\mathrm{A}$ and $|\mathcal{C}|=|\bar{\mathcal{C}}|=|\mathcal{S}|=|\bar{\mathcal{S}}|=N_\mathrm{S}$.  }
    \label{Fig1}
\end{figure*}

We consider a linear, passive, time-invariant DUT with $N_\mathrm{S}$ monomodal lumped ports. As discussed in more detail in~[Sec.~II,~\cite{del2025wireless}], we assume that waves can only enter and exit the DUT via these $N_\mathrm{S}$ ports. Our goal is to retrieve the scattering matrix $\mathbf{S}^\mathrm{DUT} \in \mathbb{C}^{N_\mathrm{S}\times N_\mathrm{S}}$ characterizing the DUT without the ability to directly inject and/or receive waves via any of its $N_\mathrm{S}$ ports. We emphasize that this assumption precludes the characterization of the DUT via the Virtual VNA technique which requires the ability to inject/receive waves via a fixed subset of DUT ports while terminating the remaining ones with a specific known TLN.

To remotely probe the DUT's scattering properties, we connect each DUT port to a distinct antenna. These $N_\mathrm{S}$ antennas are not-directly-accessible (NDA) but they couple OTA to a distinct set of $N_\mathrm{A}$ accessible antennas (away from the DUT) via which we can inject and/or receive waves. We thus probe the DUT via an OTA fixture whose characteristics are a priori unknown because they depend on the WPE (including the structural scattering of the utilized antennas). The OTA fixture is an $(N_\mathrm{A}+N_\mathrm{S})$-port system characterized by the scattering matrix $\mathbf{S}^\mathrm{OTA} \in \mathbb{C}^{(N_\mathrm{A}+N_\mathrm{S})\times (N_\mathrm{A}+N_\mathrm{S})}$ that can be characterized by the Virtual VNA technique because $N_\mathrm{A}$ of its ports are accessible. We thus insert a specific known TLN between the NDA antennas and the DUT; this TLN satisfies (almost all) the criteria of the Virtual VNA technique and additionally is capable of connecting a given NDA antenna port to the corresponding DUT port. Specifically, the TLN can terminate each NDA antenna port with three distinct individual loads or via a coupled load also connected to an adjacent NDA antenna port. The only missing ingredient with respect to the Virtual VNA requirements is the possibility to connect one accessible port and one NDA port via a coupled load. However, as discussed in~\cite{del2025wireless}, this missing ingredient only results in a sign ambiguity on certain off-diagonal entries of $\mathbf{S}^\mathrm{OTA}$ that is operationally irrelevant for the purpose of wireless multiport sensing.

In our prior work~\cite{del2025wireless}, after having characterized the OTA fixture (up to the operationally irrelevant ambiguity), we connected all NDA antenna ports to the DUT. For simplicity, we treated the connections from the NDA antennas' ports via the TLN to the DUT ports as part of the DUT in~\cite{del2025wireless}. We then conducted a \textit{single} measurement of $\mathbf{S}$ (or an off-diagonal block thereof) involving the DUT. Subsequently, we de-embedded the OTA fixture from this measurement to retrieve $\mathbf{S}^\mathrm{DUT}$. With this approach, we could only recover $\mathbf{S}^\mathrm{DUT}$ if the number of measured independent coefficients of $\mathbf{S}$, referred to as $m$ in the following, was sufficiently large with respect to the number of independent complex-valued unknowns in $\mathbf{S}^\mathrm{DUT}$, referred to as $d$ in the following. $d$ is a lower bound for $m$ but we experimentally observed in~\cite{del2025wireless} that $m$ should clearly exceed $d$ to ensure the identifiability of $\mathbf{S}^\mathrm{DUT}$, presumably due to the vulnerability of the non-linear operations to inaccuracies and noise.

Here, we deviate from the approach taken in~\cite{del2025wireless} after the sufficiently unambiguous characterization of the OTA fixture. We exploit the TLN's reconfigurability to take \textit{multiple} measurements of $\mathbf{S}$ (or an off-diagonal block thereof) involving the DUT. We refer to the number of measurements involving the DUT as $p$ in the following. In each of the $p$ measurements, we configure the TLN to connect a randomly chosen subset of the NDA antennas' ports to the corresponding DUT ports while terminating the other NDA antennas' ports with loads that we choose randomly among the TLN's loads. Thereby, we achieve measurement diversity that can compensate a low value of $m$. In the most extreme case that we demonstrate experimentally below, we can estimate $\mathbf{S}^\mathrm{DUT}$ (with $d=N_\mathrm{S}(N_\mathrm{S}+1)/2 > 1$ for a generic, reciprocal multi-port DUT) purely based on measurements of a single transmission coefficient (i.e., $m=1$) across $p$ fixture realizations.

To formalize our approach, we introduce the TLN's scattering matrix $\mathbf{S}^\mathrm{TLN}\in\mathbb{C}^{2N_\mathrm{S}\times 2 N_\mathrm{S}}$. While $\mathbf{S}^\mathrm{OTA}$ and $\mathbf{S}^\mathrm{DUT}$ are static, $\mathbf{S}^\mathrm{TLN}$ depends on how we configure the TLN. Altogether, our setup constitutes a chain cascade of $\mathbf{S}^\mathrm{OTA}$, $\mathbf{S}^\mathrm{TLN}$ and $\mathbf{S}^\mathrm{DUT}$, as shown on the left in Fig.~\ref{Fig1}. The connection between $\mathbf{S}^\mathrm{OTA}$ and $\mathbf{S}^\mathrm{TLN}$ constitutes a programmable fixture (PF) via which we probe $\mathbf{S}^\mathrm{DUT}$. PF is characterized by the following scattering matrix:
\begin{equation}
  \mathbf{S}^\mathrm{PF} =
  \begin{bmatrix}
    \mathbf{S}^\mathrm{PF}_\mathcal{AA} & \mathbf{S}^\mathrm{PF}_\mathcal{AS}\\
    \mathbf{S}^\mathrm{PF}_\mathcal{SA} & \mathbf{S}^\mathrm{PF}_\mathcal{SS}
  \end{bmatrix}
  \in \mathbb{C}^{(N_\mathrm{A}+N_\mathrm{S})\times(N_\mathrm{A}+N_\mathrm{S})}.
  \label{eq0}
\end{equation}
where $\mathcal{A}$ and $\mathcal{S}$ denote, respectively, the sets of port indices of PF associated with the accessible antennas' ports and the DUT ports. The analytical expression for $\mathbf{S}^\mathrm{PF}$ is obtained via the well-established ``Redheffer star product'' \cite{redheffer_inequalities_1959,chu_generalized_1986,overfelt1989alternate,prod2024efficient}:
\begin{equation}
\begin{split}
& {\mathbf{S}}^{\mathrm{PF}}_{\mathcal{A}\mathcal{A}} =
\mathbf{S}^\mathrm{OTA}_{\mathcal{A}\mathcal{A}}
- \mathbf{S}^\mathrm{OTA}_{\mathcal{A}\mathcal{C}}
  \mathbf{S}^\mathrm{TLN}_{\bar{\mathcal{C}}\bar{\mathcal{C}}}
  \mathbf{X}_1
  \mathbf{S}^\mathrm{OTA}_{\mathcal{C}\mathcal{A}},\\
& {\mathbf{S}}^{\mathrm{PF}}_{\mathcal{A}\mathcal{S}} =
-\mathbf{S}^\mathrm{OTA}_{\mathcal{A}\mathcal{C}}
  \mathbf{X}_2
  \mathbf{S}^\mathrm{TLN}_{\bar{\mathcal{C}}\mathcal{S}},\\
& {\mathbf{S}}^{\mathrm{PF}}_{\mathcal{S}\mathcal{A}} =
-\mathbf{S}^\mathrm{TLN}_{\mathcal{S}\bar{\mathcal{C}}}
  \mathbf{X}_1
  \mathbf{S}^\mathrm{OTA}_{\mathcal{C}\mathcal{A}},\\
& {\mathbf{S}}^{\mathrm{PF}}_{\mathcal{S}\mathcal{S}} =
\mathbf{S}^\mathrm{TLN}_{\mathcal{S}\mathcal{S}}
- \mathbf{S}^\mathrm{TLN}_{\mathcal{S}\bar{\mathcal{C}}}
  \mathbf{S}^\mathrm{OTA}_{\mathcal{C}\mathcal{C}}
  \mathbf{X}_2
  \mathbf{S}^\mathrm{TLN}_{\bar{\mathcal{C}}\mathcal{S}},
\end{split}
\label{eq1}
\end{equation}
where $\mathcal{C}$ denotes the set of TLN-side port indices of the OTA fixture, $\bar{\mathcal{C}}$ denotes the set of OTA-side port indices of the TLN, and 
\begin{equation}
\begin{split}
 & \mathbf{X}_1 = \left( \mathbf{S}^\mathrm{OTA}_{\mathcal{C}\mathcal{C}} \mathbf{S}^\mathrm{TLN}_{\bar{\mathcal{C}}\bar{\mathcal{C}}} - \mathbf{I}_{N_\mathrm{S}} \right)^{-1}, \\
 & \mathbf{X}_2 = \left( \mathbf{S}^\mathrm{TLN}_{\bar{\mathcal{C}}\bar{\mathcal{C}}} \mathbf{S}^\mathrm{OTA}_{\mathcal{C}\mathcal{C}} - \mathbf{I}_{N_\mathrm{S}} \right)^{-1}.
\end{split}
\label{eq1aux}
\end{equation}
Furthermore, according to multi-port network theory~\cite{anderson_cascade_1966,ha1981solid,prod2024efficient}, the scattering matrix $\mathbf{S}\in\mathbb{C}^{N_\mathrm{A}\times N_\mathrm{A}}$ that we can measure at the accessible antennas' ports is given by 
\begin{equation}
    \mathbf{S} =
    \mathbf{S}^\mathrm{PF}_\mathcal{AA}
    + \mathbf{S}^\mathrm{PF}_\mathcal{AS}
      \left( \left(\mathbf{S}^\mathrm{DUT}\right)^{-1} - \mathbf{S}^\mathrm{PF}_\mathcal{SS} \right)^{-1}
      \mathbf{S}^\mathrm{PF}_\mathcal{SA}.
    \label{eq4}
\end{equation}
If the accessible antennas are partitioned into $N_\mathrm{T}$ transmitting and $N_\mathrm{R}=N_\mathrm{A}-N_\mathrm{T}$ receiving ones, we can only measure a transmission matrix $\mathbf{H}\in\mathbb{C}^{N_\mathrm{R}\times N_\mathrm{T}}$ and (\ref{eq4}) specializes to
\begin{equation}
    \mathbf{H} = \mathbf{S}_\mathcal{RT}
    = \mathbf{S}^\mathrm{PF}_\mathcal{RT}
    + \mathbf{S}^\mathrm{PF}_\mathcal{RS}
      \left( \left(\mathbf{S}^\mathrm{DUT}\right)^{-1} - \mathbf{S}^\mathrm{PF}_\mathcal{SS} \right)^{-1}
      \mathbf{S}^\mathrm{PF}_\mathcal{ST},
    \label{eq5}
\end{equation}
where $\mathcal{T}$ and $\mathcal{R}$ denote the sets of port indices associated with transmitting and receiving accessible antennas, respectively, and $\mathcal{A}=\mathcal{T}\cup\mathcal{R}$.

We emphasize that the present work does \textit{not} hinge on the details of the wave generation and detection apparatus. If an $N_\mathrm{A}$-port VNA is available, it can be directly connected to all accessible antennas and supply the measurable scattering matrix $\mathbf{S} \in \mathbb{C}^{N_\mathrm{A}\times N_\mathrm{A}}$. If a multiple-input multiple-output (MIMO) setup is used in transmission, the accessible antennas are partitioned into transmitting and receiving ones. Provided the MIMO hardware is capable of coherent transmission and reception, one can measure $\mathbf{H}$. 
We focus on such transmission setups yielding measurements of $\mathbf{H}$ in this article because they most closely resemble most conceivable wave generation and detection setups without a VNA. 
In particular, a SISO setup based on two SDRs constitutes a special case of our general MIMO formulation with $N_\mathrm{A}=2$ (i.e., one transmitting accessible antenna, and one distinct receiving accessible antenna) in which $\mathbf{H}$ collapses to a scalar. Such SDR-based SISO setups are common in RFID contexts~\cite{vena2024backscatter} and distinguished by their particularly low hardware complexity because they do not require coherent wavefront generation or reception. However, they imply $m=1$ so that it remained elusive to wirelessly characterize a multi-port DUT ($d>1$) with such setups. Our present work makes this possible, as we experimentally demonstrate  below.

To summarize, our goal is to estimate the $d$ independent parameters in $\mathbf{S}^\mathrm{DUT}$ remotely (i.e., without injecting/receiving waves via any of the DUT ports) based on measurements of $m<d$ (ideally, $m=1$) independent scattering coefficients at accessible antenna ports for $p$ TLN realizations.

From the perspective of de-embedding, our method overcomes the impossibility of de-embedding a known \textit{static} OTA fixture with too few accessible ports. By combining this underdetermined OTA fixture with the TLN, we obtain a known \textit{programmable} fixture with the same port count as the OTA fixture. Thus, a single known realization of the PF is also impossible to de-embed if the OTA fixture has too few accessible ports. However, the ability to measure the DUT with different known realizations of the PF allows us to jointly de-embed an ensemble of PF realizations. We refer to this approach as ``multiplexed de-embedding'' because we multiplex DUT information across different PF realizations.

\section{Comparison with Compressive Sensing}
\label{sec_CScomparison}

Our idea to estimate $d$ unknowns based on $m<d$ measurements across $p$ realizations of measurement diversity is inspired by compressive sensing techniques, in particular  single-pixel cameras~\cite{duarte2008single}. The embodiment of the latter that most closely resembles our present work is the use of dynamic metasurface antennas (DMAs) for single-frequency computational microwave imaging~\cite{sleasman2015dynamic,li2016transmission,sleasman2016microwave,sleasman2017single,sleasman2020implementation}. Nonetheless, there are important conceptual differences. 

\textit{First}, computational imaging typically discretizes the scene into voxels under a local approximation, meaning that the reflection from a given voxel is assumed to only depend on the field impinging on that voxel but not on the fields impinging on other voxels. In our problem, the DUT ports are naturally discrete but the reflected signal exiting any given DUT port generally depends on the signals entering all DUT ports. A local approximation would only be valid in our problem if we specialized our problem formulation to considering DUTs with diagonal scattering matrices. 

\textit{Second}, most works on computational imaging rely on a first Born approximation that neglects multiple scattering between different voxels in the scene to be imaged as well as between the scene and the transceiver side. Given these assumptions, the mapping from scene reflectivity to the measurements is \textit{linear} by construction. In our problem, irrespective of whether the DUT is diagonal or not, there is generally mutual coupling between the DUT ports via the PF. Waves also generally scatter multiple times between the accessible ports and DUT ports within the PF, especially if the WPE of the OTA fixture is not simply free space. As a result, our forward model in (\ref{eq4}) mapping the unknown $\mathbf{S}^\mathrm{DUT}$ to the measurement $\mathbf{S}$, or a block thereof as in (\ref{eq5}), is inherently \textit{non-linear}. 

\textit{Third}, compressive sensing algorithms rely by construction on an assumption of sparsity of the sought-after information.
Our goal in this paper is to achieve the identifiability of the unknowns, without insisting on achieving this with fewer measurements than unknowns. Consequently, we do \textit{not} assume that there exists a sparse representation of $\mathbf{S}^\mathrm{DUT}$. However, future work can explore our problem statement in light of existing theoretical works on compressive sensing with non-linear observations~\cite{Blumensath2013}.

\textit{Fourth}, one can argue that the illumination diversity on which computational imaging relies is fundamentally different from the PF diversity on which our work relies. However, we do not insist on this argument because it is subjective. Indeed, one can also argue that programming the fixture is a way of generating diverse illuminations. 

Finally, we point out that computational imaging typically aims to \textit{qualitatively} reconstruct the scene reflectivity (under conditions in which the stated assumptions approximately apply), which is sufficient for envisioned applications such as security screening. In contrast, our goal here is to \textit{quantitatively} reconstruct the DUT's scattering matrix fully and unambiguously.

\section{Method}
\label{sec_Method}

\begin{figure}[b]
    \centering
    \includegraphics[width=0.9\columnwidth]{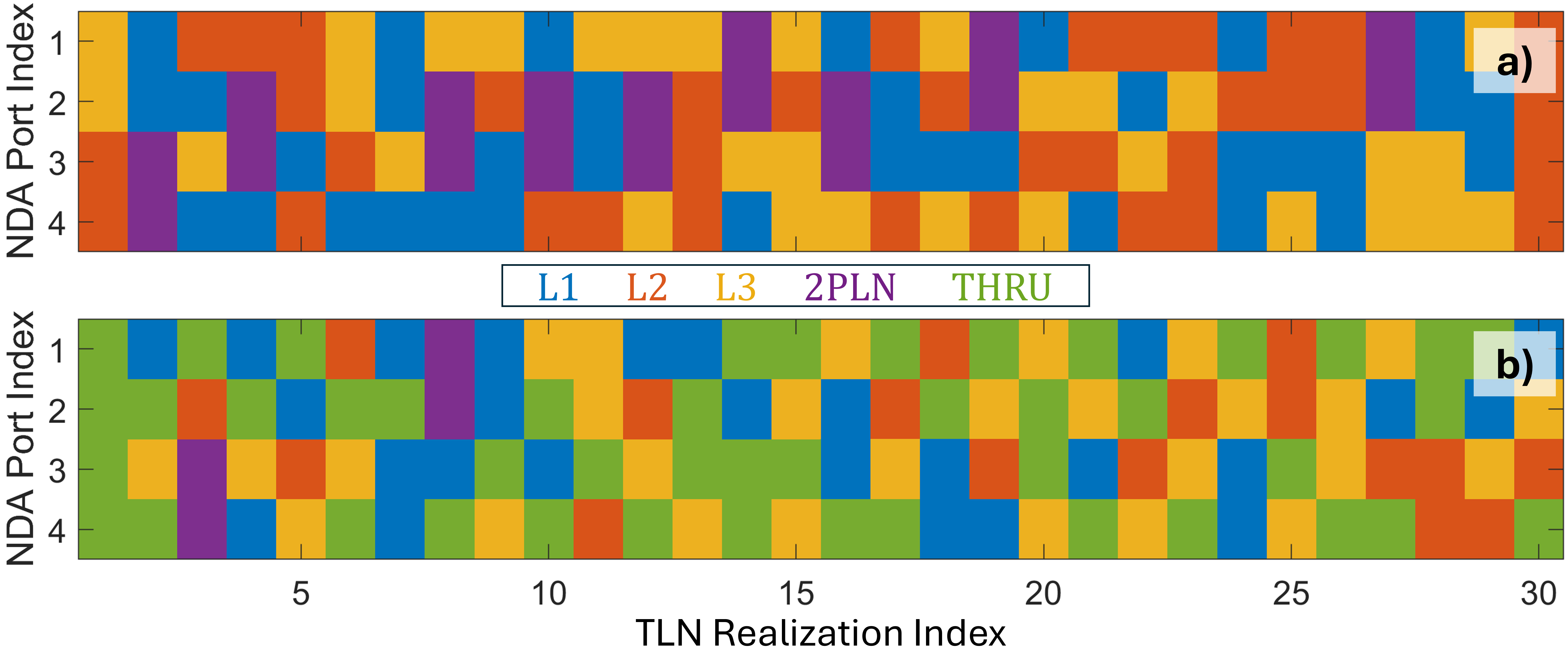}
    \caption{Utilized TLN configurations for (a) Step 1 (Virtual-VNA-based OTA fixture characterization), and (b) Step 2 (DUT measurements with different PF realizations).  }
    \label{Fig2}
\end{figure}

We proceed in three steps. The first step is the same Virtual-VNA-based characterization of the OTA fixture as in~\cite{del2025wireless}. In short, we measure $\mathbf{H}$ for a series of known TLN configurations. None of these TLN configurations establishes a connection to the DUT, and we ensure that over the course of the series of configurations we use each possible termination for each port at least once. We also avoid repeating the same TLN configuration; in total, there are $\sum_{r=0}^{\left\lfloor N_\mathrm{S}/2 \right\rfloor} \binom{N_\mathrm{S} - r}{\,r\,}\,3^{\,N_\mathrm{S} - 2r}$ distinct TLN configurations. An example series of TLN configurations for characterizing the OTA fixture is shown in Fig.~\ref{Fig2}a. More details on  this first step can be found in~\cite{del2025wireless}, as well as earlier Virtual-VNA literature. 

After this first step, we deviate from~\cite{del2025wireless}. We conduct a series of $p$ measurements of $\mathbf{H}$, each corresponding to a distinct TLN configuration. This time, each TLN configuration contains at least one connection from an NDA antenna to the corresponding DUT port. In total, there are $\sum_{r=0}^{\left\lfloor N_\mathrm{S}/2\right\rfloor}
\binom{N_\mathrm{S} - r}{\,r\,}\,\Big(4^{\,N_\mathrm{S}-2r}-3^{\,N_\mathrm{S}-2r}\Big)$ distinct TLN configurations satisfying this criterion. Moreover, we choose to connect all NDA antennas to their DUT ports in the first configuration. 
An example of such a series of TLN configurations is shown in Fig.~\ref{Fig2}b. To be clear, only the very first of these configurations was used in~\cite{del2025wireless} (implying $p=1$), which precluded the remote characterization of many-port DUTs with a setup limited to a low value of $m$.

The third step concerns the de-embedding of $\mathbf{S}^\mathrm{DUT}$ from the measurements of $m$ entries of $\mathbf{H}$ across $p$ PF realizations. The $p$ utilized realizations of $\mathbf{S}^\mathrm{TLN}$ are known, and $\mathbf{S}^\mathrm{OTA}$ is known (sufficiently unambiguously) from Step 1. Thus, given~(\ref{eq1}), $\mathbf{S}^\mathrm{PF}$ is known sufficiently unambiguously for the $p$ utilized PF realizations. We emphasize that this is an unconventional ``multiplexed de-embedding'' problem when $p>1$, as mentioned earlier. We tackle this problem with gradient descent. Specifically, we parametrize the reciprocal DUT as a symmetric matrix via its $d$ upper-triangular entries, i.e., our estimate of $\mathbf{S}^\mathrm{DUT}$ is $\Sym(\bm\theta) $, where $\bm\theta \in \mathbb{C}^{d}$; see~(\ref{eq12}) for details. Then, we define a cost $\mathcal{L}$ that quantifies the relative element-wise $\ell_1$ misfit between measurement (meas) and prediction (pred) of the observed scattering coefficient(s) across all PF realizations:
\begin{equation}
\mathcal{L}(\bm\theta) =
\frac{\sum_{r=1}^p \big\|(\mathbf{H}^{\text{pred}}_r(\bm\theta)-\mathbf{S}^{\mathrm{PF}(r)}_{\mathcal{RT}})
-(\mathbf{H}^{\text{meas}}_r-\mathbf{S}^{\mathrm{PF}(r)}_{\mathcal{RT}})\big\|_1}
{\sum_{r=1}^p \big\|\mathbf{H}^{\text{meas}}_r-\mathbf{S}^{\mathrm{PF}(r)}_{\mathcal{RT}}\big\|_1},
\label{eq6}
\end{equation}
where the index $r$ identifies the $r$th PF realization.
We minimize $\mathcal{L}$ by gradient descent (Adam) with a decaying step size, where we obtain the gradients with TensorFlow's automatic differentiation through the required matrix solves and multiplications of our forward model in (\ref{eq5}). We have not extensively explored different definitions of $\mathcal{L}$ because the one in (\ref{eq6}) yielded good results, as seen in Fig.~\ref{Fig_MainResults} below; we defer explorations of alternative definitions of $\mathcal{L}$ to future work.

\section{Experimental Validation}
\label{sec_ExpVal}

We describe our experimental setup and procedure in Sec.~\ref{subsec_expSetupProc}. Then, we analyze the experimentally achieved PF diversity in Sec.~\ref{subsec_AnalysisPFdiversity}. Finally, we examine the resulting estimates of $\mathbf{S}^\mathrm{DUT}$ in Sec.~\ref{subsec_expResults}.

\subsection{Experimental Setup and Procedure}
\label{subsec_expSetupProc}

We conduct our experiments at 2.45~GHz based on the setup shown in the middle in Fig.~\ref{Fig1}. Our DUT is a four-port, reciprocal circuit composed of a complicated mesh of transmission lines. We have no prior knowledge about $\mathbf{S}^\mathrm{DUT}$ besides its symmetry due to reciprocity. Our OTA fixture features rich scattering inside a reverberation chamber so that we cannot have any prior knowledge about $\mathbf{S}^\mathrm{OTA}$ other than reciprocity. Our setup comprises three arrays of four standard WiFi antennas (ANT-24G-HL90-SMA) with half-wavelength spacing; one is connected to the DUT via the TLN, one is the transmit array, and one is the receive array. The latter two are cross-polarized to minimize the line-of-sight contribution.

\begin{figure}
    \centering
    \includegraphics[width=\columnwidth]{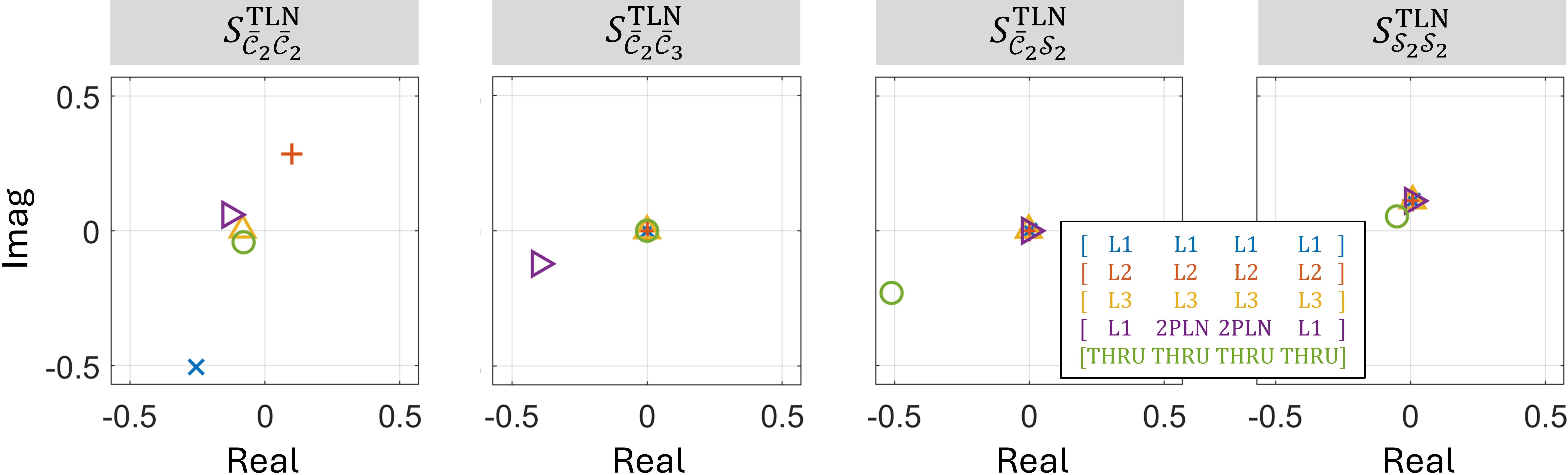}
    \caption{Representative TLN scattering coefficients at 2.45~GHz for five representative TLN configurations. Left to right: reflection coefficient of an OTA-side TLN port, transmission coefficient between two adjacent OTA-side TLN ports, transmission coefficient between an OTA-side TLN port and the corresponding DUT-side TLN port, reflection coefficient of a DUT-side TLN port.}
    \label{Fig3}
\end{figure}

Our TLN is composed of four SP8T switches (HMC321ALP4E, Analog Devices), as seen in the middle panel in Fig.~\ref{Fig1} (the fifth switch in our setup on the right side of the TLN is not used). The input port of each switch is connected to an NDA antenna, and one output port is connected to a DUT port. Three other output ports are connected to individual loads. Two further output ports are connected via coupled loads (simple connectors) to adjacent switches. The three individual loads can be chosen arbitrarily as long as they are distinct; for simplicity we use connectorized open-circuit, short-circuit, and matched loads. However, we emphasize that due to propagation in the switch, the corresponding individual loads seen from the TLN's OTA-side ports are \textit{not} calibration-standard loads, as seen in the leftmost panel in Fig.~\ref{Fig3}.

In a calibration step, we use an eight-port VNA (two cascaded Keysight P5024B 4-port VNAs) to measure $\mathbf{S}^\mathrm{TLN}\in\mathbb{C}^{8\times 8}$ for the 30 random TLN configurations shown in Fig.~\ref{Fig2}b. Then, using the all-THRU configuration of the TLN, we measure the $4\times 4$ scattering matrix resulting from the connection between TLN and DUT. We only use this measurement to obtain the ground-truth $\mathbf{S}^\mathrm{DUT}$ by de-embedding the known TLN. (This procedure is easier than directly measuring $\mathbf{S}^\mathrm{DUT}$ for practical reasons related to available connectors.) For our main measurements, we connect the OTA-side TLN ports via coaxial cables to the NDA antennas. We connect our eight-port VNA to the eight accessible antennas' ports and measure $\mathbf{S}\in\mathbb{C}^{8\times 8}$. For any choice of transmitting and receiving antennas, we can extract the corresponding $\mathbf{H}$ from the measured $\mathbf{S}$. If $N_\mathrm{A}<8$, we thus assume that unused accessible antennas are terminated in matched loads. 

We conduct our experiments under well-controlled conditions: The signal-to-noise ratio (SNR) is 62.3~dB (estimated based on repeated measurements in quick succession of $\mathbf{S}$ with all switches in a reference configuration) and the stability is 46.5~dB (estimated analogously to the SNR, but based on measurements taken intermittently over the course of our experiment).
We make every possible effort to minimize mechanical manipulations of the coaxial cables; where possible, they are taped to the table to maximize stability (see Fig.~\ref{Fig1}). However, two manual interventions are necessary (first, to connect the TLN DUT-side ports to the DUT; second, to connect the TLN OTA-side ports to the NDA antennas) which inevitably introduce some imperfections.

\subsection{Analysis of PF Diversity}
\label{subsec_AnalysisPFdiversity}

Our goal in this subsection is to examine the measurement diversity originating from PF diversity. For conventional computational microwave imagers with a linear forward model, one simply examines the singular-value (SV) distribution of the sensing matrix that maps scene reflectivity to measurements. For instance, using more random DMA configurations dampens the decay of the SV spectrum in~\cite{sleasman2016microwave,sleasman2020implementation}, which evidences that using more random DMA configurations improves the measurement diversity. Optimized rather than random configurations can even achieve an essentially flat SV spectrum~\cite{del2020optimal}, indicating close-to-ideal measurement diversity.
In our present problem, however, we are confronted with the non-linear forward operator defined in~(\ref{eq5}). Thus, we cannot straightforwardly define a sensing matrix and compute its SV spectrum to examine how the latter depends on $p$. 

To formalize our PF-diversity analysis, let us denote by $\mathbf{y}\in\mathbb{C}^{mp}$ our vector of $mp$ independent measurements ($m$ independent scattering coefficients for each of $p$ PF realizations). Our measurements depend via a non-linear function $f$ on our chosen set of PF configurations (captured by some $\mathbf{C}\in\mathbb{R}^{p\times N_\mathrm{S}}$ that is illustrated in Fig.~\ref{Fig2}b) and the DUT itself (whose $\mathbf{S}^\mathrm{DUT}$ is characterized by $\bm\theta$):
\begin{equation}
 \mathbf{y} = f(\mathbf{C},\bm\theta).   
  \label{eq_7}
\end{equation}
Following~\cite{Blumensath2013}, we linearize $f$ around a reference $\bm{\theta}_0$:
\begin{equation}
  f(\mathbf{C},\bm{\theta})
  \;\approx\;
  \mathbf{y}_0 \;+\; \mathbf{J}(\mathbf{C},\bm{\theta}_0)\,\delta\bm{\theta},
  \label{eq_8}
\end{equation}
where $\mathbf{y}_0 \!=\! f(\mathbf{C},\bm{\theta}_0)$, $\mathbf{J}(\mathbf{C},\bm{\theta}_0) = \left.\frac{\partial f(\mathbf{C},\bm{\theta})}{\partial \bm{\theta}}\right|_{\bm{\theta}_0} \in \mathbb{C}^{mp\times d}$ , and $\delta\bm{\theta}=\bm{\theta}-\bm{\theta}_0$. 
From~(\ref{eq_8}), we see that the Jacobian $\mathbf{J}(\mathbf{C},\bm{\theta}_0)$ plays the role of the sensing matrix in the vicinity of $\bm{\theta}_0$ because it tells us how small changes in the DUT parameters (from $\bm{\theta}_0$ to $\bm{\theta}$) propagate to our measurements $\mathbf{y}$. To see whether PF-diversity improves the DUT's identifiability, we thus examine the SV spectrum of the Jacobian $\mathbf{J}(\mathbf{C},\bm{\theta}_0)$ as a function of $p$. To quantify the flatness of the Jacobian's SV spectrum, we compute its effective rank~\cite{roy2007effective}
\begin{equation}
    R(\mathbf{J}) = \mathrm{exp} \left(-\sum_{k=1}^{\tilde{N}} \hat{\iota}_k \mathrm{ln}(\hat{\iota}_k)\right),
\label{eq_9}
\end{equation}
where $\tilde{N}=\mathrm{min}(mp,d)$, $\hat{\iota}_k = \iota_k / \sum_{l=1}^{\tilde{N}} {\iota_l} $, and $\iota_k$ is the $k$th SV (in descending order) of $\mathbf{J}$. It follows that $1 \leq R \leq \tilde{N}\leq d$.

\begin{figure}
    \centering
    \includegraphics[width=0.75\columnwidth]{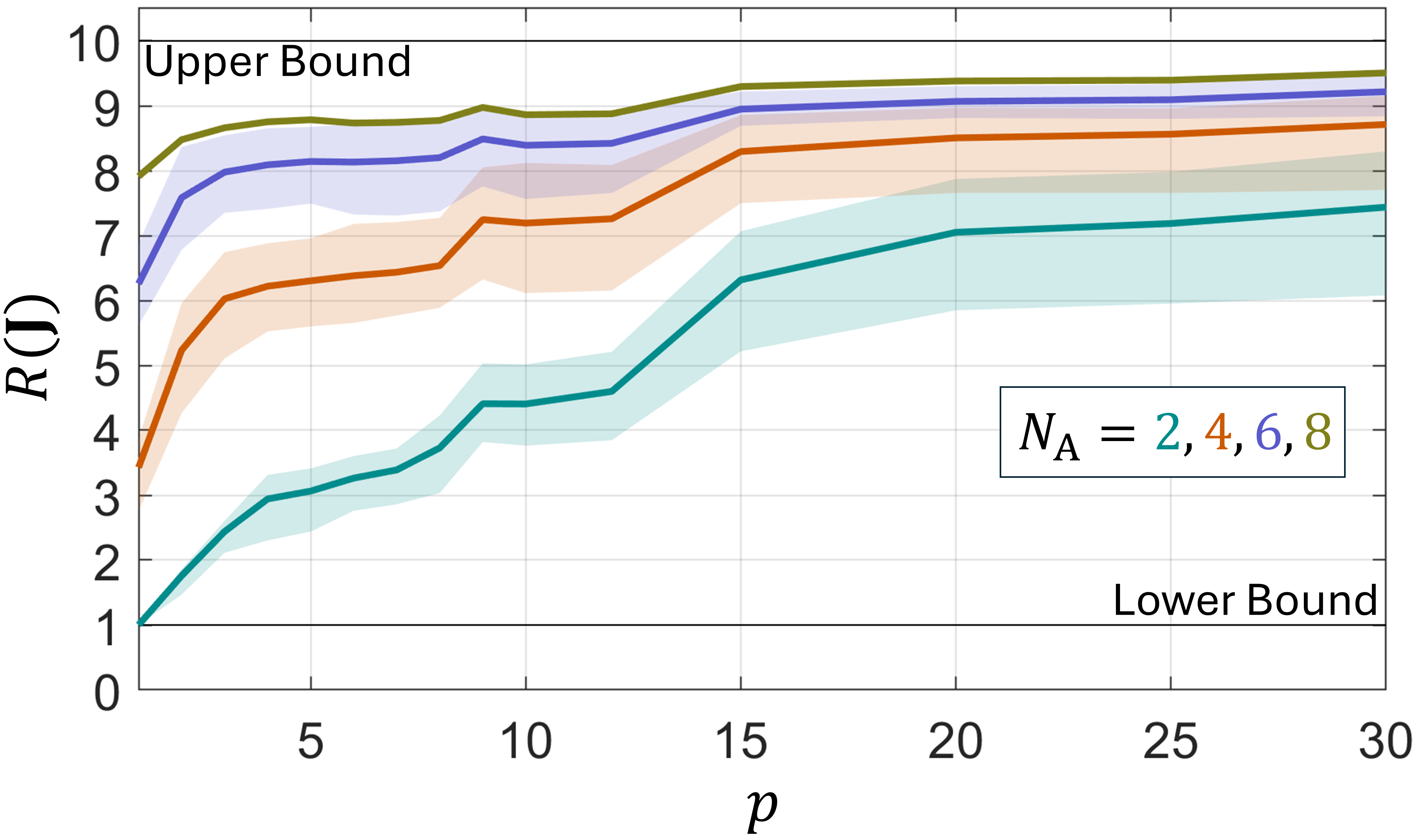}
    \caption{Dependence of $R(\mathbf{J})$ (see~(\ref{eq_9})~for definition) on $p$ for the four considered values of $N_\mathrm{A}$ (see legend for color code). The continuous line and shade represent the mean and the span between minimum and maximum, respectively, across all possible choices of TX and RX.}
    \label{Fig4}
\end{figure}

Given our forward model in~(\ref{eq5}), our estimate of $\mathbf{S}^\mathrm{OTA}$, and our knowledge of $\mathbf{S}^\mathrm{TLN}$ for the $p$ TLN realizations in $\mathbf{C}$, we can evaluate $\mathbf{J}$ in closed form for a given $\mathbf{S}^\mathrm{DUT}$. The analytical expression is provided in the Appendix. We have computed $R(\mathbf{J})$ for each considered value of $p$ and each considered value of $m$ (separately for each possible choice of TX and RX), using the $\bm{\theta}_0$ that corresponds to our ground-truth measurement of $\mathbf{S}^\mathrm{DUT}$. A summary of our results is displayed in Fig.~\ref{Fig4}. We also repeated the same procedure for choices of $\bm{\theta}_0$ corresponding to random numerical realizations of $\mathbf{S}^\mathrm{DUT}$ (subject to passivity and reciprocity); the results were extremely similar to those displayed in Fig.~\ref{Fig4}.

We observe in Fig.~\ref{Fig4} that $R(\mathbf{J})$ increases with both $m$ and $p$. The increase of $R(\mathbf{J})$ with $p$ confirms that PF diversity improves the identifiability of $\mathbf{S}^\mathrm{DUT}$, which we validate explicitly in the next subsection. 
For $m=16$ ($N_\mathrm{A}=8$), $R(\mathbf{J})$ is around 7.9 for $p=1$ which is already quite close to its upper bound of 10. Indeed, with $m=16$, the considered problem is well-posed even for $p=1$; consequently, larger values of $p$ do not bring about a significant increase in $R(\mathbf{J})$, reaching 9.5 at $p=30$. Given the random nature of our PF realizations, we can expect $\mathbf{J}$ to be at best pseudo-random, meaning that $R(\mathbf{J})$ would reach its upper bound only as $p$ becomes very large. However, there is an upper bound on the number of distinct PF realizations, as mentioned earlier. 
For $m=9$ ($N_\mathrm{A}=6$), $R(\mathbf{J})$ rises from 6.3 with $p=1$ to 7.6 with $p=2$ and finally to 9.2 with $p=30$. For low values of $p$, the benefits of increasing $p$ are thus more substantial than if $p$ is already large.
A similar but more marked trend is observed for $m=4$ ($N_\mathrm{A}=4$), where $R(\mathbf{J})$ rises from 3.4 with $p=1$ to 6.0 with $p=3$ and finally to 8.7 with $p=30$.
For $m=1$ ($N_\mathrm{A}=2$), $R(\mathbf{J})$ starts out at its lower bound of unity for $p=1$ and rises all the way to 7.4 at $p=30$. This remarkably evidences the benefit of PF diversity.

We also observe in Fig.~\ref{Fig4} that none of the curves has saturated at $p=30$, meaning that further increases in $R(\mathbf{J})$ for $p>30$ can be expected, especially for lower values of $m$. Altogether, the results in Fig.~\ref{Fig4} give us confidence that PF diversity is a viable path to achieving the identifiability of $\mathbf{S}^\mathrm{DUT}$ with very low values of $m$. Our results in the next subsection confirm this.

\subsection{Analysis of $\mathbf{S}^\mathrm{DUT}$ Estimation}
\label{subsec_expResults}

\begin{figure*}
    \centering
    \includegraphics[width=1.5\columnwidth]{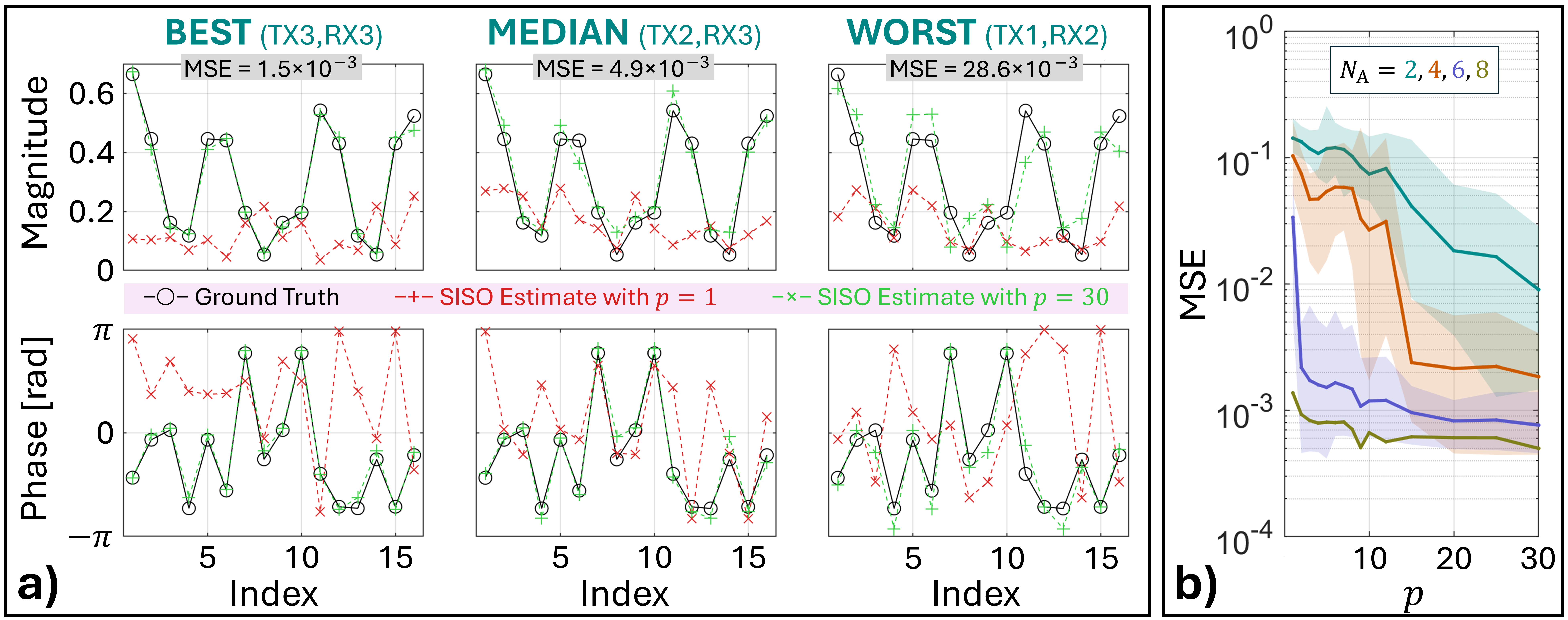}
    \caption{(a) Representative remote characterizations of a 4-port reciprocal DUT based on SISO measurements ($N_\mathrm{A}=2$) for $p=1$ (red) and $p=30$ (green), benchmarked against the ground truth (black). The 16 entries of $\mathbf{S}^\mathrm{DUT}$ are vectorized and indexed along the horizontal axis. (b) Systematic analysis of the MSE (based on measurements of $\mathbf{H}$) as a function of $p$ for four values of $N_\mathrm{A}$ (see legend for color code). The continuous lines and shades represent, respectively, the mean and the span from minimum to maximum, across all 16 possible choices of TX and RX in our setup.}
    \label{Fig_MainResults}
\end{figure*}

We begin by qualitatively examining the fidelity of representative estimates of $\mathbf{S}^\mathrm{DUT}$ based on measurements of a single transmission coefficient (i.e., $m=1$) across $p$ PF realizations. This corresponds to an extremely low-cost setup that emulates the common SDR-based SISO setup used in the realm of RFID~\cite{vena2024backscatter}. Given our experimental data acquired with the setup in Fig.~\ref{Fig1}, we have 16 different choices of which transmission coefficient to use. Each corresponds to a different PF and yields thus a somewhat different accuracy. We consider three representative examples (best, median, worst) in Fig.~\ref{Fig_MainResults}a. In all three cases, it is apparent that $\mathbf{S}^\mathrm{DUT}$ is not identifiable with $p=1$. This was expected because based on one complex-valued measurement (one transmission coefficient for one PF realization) there is no hope to estimate 10 independent unknowns. However, by measuring the transmission coefficient across a moderate amount ($p=30$) of PF realizations, $\mathbf{S}^\mathrm{DUT}$  becomes identifiable in all three cases. Even in the worst case, ground truth and estimate of $\mathbf{S}^\mathrm{DUT}$ display strong correlations. Consequently, our method of multiplexing across PF realizations enables remote \textit{multi}-port DUT characterizations with a low-cost SDR-based \textit{SISO} setup like the one in~\cite{vena2024backscatter}.

Next, we examine quantitatively how the mean-squared error (MSE) of the estimate of $\mathbf{S}^\mathrm{DUT}$ scales with $m$ and $p$. Besides the aforementioned SISO case (corresponding to $N_\mathrm{A}=2$), we consider $2\times 2$ MIMO, $3\times 3$ MIMO, and $4\times 4$ MIMO (corresponding, respectively, to $N_\mathrm{A}=4$, $N_\mathrm{A}=6$, and $N_\mathrm{A}=8$). In each case, we examine all possible choices of TX and RX among the available ones. The averages across all possible choices is plotted as continuous lines in Fig.~\ref{Fig_MainResults}b, while the spans between the best and worst case choices are indicated with shaded areas. For $4\times 4 $ MIMO, $m=16$ so that $\mathbf{S}^\mathrm{DUT}$ is identifiable even with $p=1$; using larger values of $p$ thus only yields small improvements. For $3\times 3$ MIMO, $m=9$ which is below the lower bound of $d=10$. Consequently, the MSE with $p=1$ is high and $\mathbf{S}^\mathrm{DUT}$ is not identifiable. However, the MSE sharply drops for $p>1$, and already with $p=2$ we reach identifiability. For $2\times 2 $ MIMO, $m=4$ is clearly below $d=10$ so that $\mathbf{S}^\mathrm{DUT}$ is not identifiable until $p$ exceeds 12. Again, a sharp drop in MSE is apparent in Fig.~\ref{Fig_MainResults}b around this threshold. For the SISO case with $m=1$, the MSE decreases more smoothly with $p$. We are not aware of an explanation for this smoothness (as opposed to a sharp drop) in the $m=1$ case; this smooth trend suggests that even lower SISO-based MSEs are accessible with $p>30$ in our experiment. Our visual inspection of Fig.~\ref{Fig_MainResults}a suggests that even with $p=30$ we have reached a reasonably high degree of identifiability for the SISO case.

\section{Conclusion}
\label{sec_conclusion}

To summarize, we paved the way toward wireless multi-port sensing with low-complexity hardware. We uncoupled the number of required independent transmission coefficients at accessible antennas from the number of unknowns in the DUT's scattering characteristics. We achieved this by multiplexing the DUT information across different configurations of the TLN, and hence of the PF via which we measure the DUT. Our approach does not require any additional hardware because the TLN is required to characterize the OTA fixture in any case. Our most striking result is the characterization of a reciprocal 4-port DUT (10 complex-valued unknowns) based on remote measurements of a single transmission coefficient across 30 PF realizations. Thereby, we make wireless multiport sensing accessible to low-cost SDR-based SISO setups. We also investigated more systematically the PF diversity and achieved accuracy as a function of the number of measured transmission coefficients and utilized PF realizations.

We furthermore clarified that our technique fundamentally differs from conventional compressive sensing schemes such as single-pixel computational imagers because we cannot rely on a local approximation of the DUT's scattering properties and our observations are inherently non-linear due to mutual coupling.

Looking forward, we envision four conceptual future developments. 
\textit{First}, the PF diversity can be further enhanced by integrating an RIS into the OTA fixture. A conventional RIS with 1-bit programmable RIS elements is sufficient because the ensuing ambiguities in the OTA characteristics are irrelevant for the purpose of wireless multiport sensing.
\textit{Second}, novel architectures of the TLN can be explored that enable wireless multi-port sensing with fewer NDA antennas than there are DUT ports.
\textit{Third}, the TLN configurations can be optimized based on prior knowledge, building on recent advances from compressive to learned sensing~\cite{saigre2022intelligent}. Specifically, one can optimize the TLN configuration (i) to maximize $R(\mathbf{J})$ for arbitrary DUTs (generalizing~\cite{del2020optimal}), (ii) to maximize $R(\mathbf{J})$ for specific classes of DUTs (generalizing~\cite{liang2015reconfigurable,li2019machine}), or (iii) to optimize a task-specific performance metric that hinges on knowledge of the DUT characteristics (generalizing~\cite{del2020learned,qian2022noise}).
\textit{Fourth}, the hardware complexity can be further reduced by accommodating a limitation to non-coherent detection, which we expect to be feasible given that coherent detection is dispensable for the Virtual VNA technique~\cite{del2024virtual,del2024virtual2p0,del2025virtual3p1}.

\appendix

In this Appendix, we derive $\mathbf{J}$ in closed form. 

We begin by deriving the Jacobian $\mathbf{J}^{(r)}\in\mathbb{C}^{m\times d}$ for the transmission coefficient(s) measured with the $r$th PF configuration. For notational ease, let us define $\mathbf{A}=\mathbf{S}^{\mathrm{PF}(r)}_{\mathcal{RS}}$, $\mathbf{B}=\mathbf{S}^{\mathrm{PF}(r)}_{\mathcal{ST}}$, $\mathbf{D}=\mathbf{S}^{\mathrm{PF}(r)}_{\mathcal{SS}}$, and $\mathbf{G}\left(\mathbf{S}^\mathrm{DUT}\right)=\left(\left(\mathbf{S}^\mathrm{DUT}\right)^{-1}-\mathbf{D}\right)^{-1}$. Consequently, $\mathbf{H}^{(r)}=\mathbf{S}^{\mathrm{PF}(r)}_{\mathcal{RT}}+\mathbf{A}\mathbf{G}\mathbf{B}$. 

Using the property $\mathrm{d}\mathbf{X}^{-1}=-\mathbf{X}^{-1}(\mathrm{d}\mathbf{X})\mathbf{X}^{-1}$~\cite{hjorungnes2007complex}, we find
\begin{equation}
\begin{aligned}
\mathrm{d}\mathbf{G}
&=-\mathbf{G}\,\mathrm{d}\big((\mathbf{S}^{\mathrm{DUT}})^{-1}-\mathbf{D}\big)\,\mathbf{G}\\
&=\mathbf{G}\,(\mathbf{S}^{\mathrm{DUT}})^{-1}\,(\mathrm{d}\mathbf{S}^{\mathrm{DUT}})\,(\mathbf{S}^{\mathrm{DUT}})^{-1}\mathbf{G}.
\end{aligned}
\end{equation}
Hence,
\begin{equation}
\mathrm{d}\mathbf{H}^{(r)}=\mathbf{A}\,\mathbf{G}\,(\mathbf{S}^{\mathrm{DUT}})^{-1}\,(\mathrm{d}\mathbf{S}^{\mathrm{DUT}})\,(\mathbf{S}^{\mathrm{DUT}})^{-1}\mathbf{G}\,\mathbf{B}.
\label{eq11}
\end{equation}

Because of reciprocity, $ \mathbf{S}^\mathrm{DUT} $ is parametrized by its upper-triangular entries, which are summarized in $\bm\theta$, as mentioned earlier. Specifically, 
\begin{equation}
\mathbf{S}^{\mathrm{DUT}} \;=\; \Sym(\bm\theta) \;=\;\sum_{k=1}^{d} \theta_k\, \mathbf{E}_k,
\label{eq12}
\end{equation}
where
\begin{equation}
\mathbf{E}_k \;=\;
\begin{cases}
\mathbf{e}_{i_k}\,\mathbf{e}_{i_k}^{\top}, & i_k = j_k,\\[4pt]
\mathbf{e}_{i_k}\,\mathbf{e}_{j_k}^{\top} + \mathbf{e}_{j_k}\,\mathbf{e}_{i_k}^{\top}, & i_k < j_k,
\end{cases}
\label{eq13}
\end{equation}
$\mathbf{e}_{i_k}\in\mathbb{C}^{N_\mathrm{S}}$ denotes the $i_k$th canonical basis vector, and $\{(i_k,j_k)\}_{k=1}^d$ enumerates the upper-triangular index set (e.g., column-wise: $j=1,\dots,N_\mathrm{S}$ and, for each $j$, $i=1,\dots,j$).

Substituting $\mathrm{d}\mathbf{S}^{\mathrm{DUT}}=\sum_{k=1}^d \mathbf{E}_k\,\mathrm{d}\theta_k$ into (\ref{eq11}) yields the columns of $\mathbf{J}^{(r)}$. Specifically,  $\mathbf{J}_k^{(r)}\in \mathbb{C}^{m}$ is the $k$th column of $\mathbf{J}^{(r)}$ and given by
\begin{equation}
    \begin{aligned}
    \mathbf{J}_k^{(r)} &= \mathrm{vec} \left(\frac{\partial \mathbf{H}^{(r)}}{\partial \theta_k}\right)
    \\&=\mathrm{vec} \left(\mathbf{A}\,\mathbf{G}\,(\mathbf{S}^{\mathrm{DUT}})^{-1}\,\mathbf{E}_k\,(\mathbf{S}^{\mathrm{DUT}})^{-1}\,\mathbf{G}\,\mathbf{B}\right),
    \end{aligned}
\end{equation}
where $\mathrm{vec}(\cdot)$ denotes vectorization.

Finally, we obtain $\mathbf{J}$ by vertically concatenating all $p$ realizations of $\mathbf{J}^{(r)}$:
\begin{equation}
\mathbf{J}
=\big[\,(\mathbf{J}^{(1)})^\top \;\cdots\; (\mathbf{J}^{(p)})^\top\,\big]^\top .
\end{equation}

\section*{Acknowledgment}
The author acknowledges stimulating discussions with N.~Barbot and A.~Vena.

\bibliographystyle{IEEEtran}

\providecommand{\noopsort}[1]{}\providecommand{\singleletter}[1]{#1}%

\end{document}